\documentclass[preprint]{ptephy_v1}%%%%%% to generate preprint number with ptep logo

\preprintnumber{RIKEN-iTHEMS-Report-23} %%% %%% Insert preprint number here
\usepackage{hyperref}
\usepackage{amsmath}% for dealing with mathematics,
\usepackage{graphics}% for dealing with figures.
\usepackage{bm}
\usepackage{color}

\def\lesssim{\ \raise.3ex\hbox{$<$}\kern-0.8em\lower.7ex\hbox{$\sim$}\ }
\def\gesim{\ \raise.3ex\hbox{$>$}\kern-0.8em\lower.7ex\hbox{$\sim$}\ }

\begin{document}

\title{Finite-range effect in the two-dimensional density-induced BCS-BEC crossover}

%%%% To generate auto affiliation numbers please use \author{}\affil{} command

\author[1,2]{Hikaru Sakakibara}
\author[1,3]{Hiroyuki Tajima}
\author[1,2]{Haozhao Liang}
\affil[1]{Department of Physics, Graduate School of Science, The University of Tokyo, Tokyo 113-0033, Japan}
\affil[2]{Interdisciplinary Theoretical and Mathematical Sciences Program (iTHEMS), RIKEN, Wako, Saitama 351-0198, Japan}
\affil[3]{RIKEN Nishina Center, Wako, Saitama 351-0198, Japan}

%%% To include the collaborator name... Please use the command "\collaborator"
%%% For example: \collaborator{ATLAS Collaboration}

\begin{abstract}%
We theoretically investigate the Bardeen-Cooper-Schrieffer (BCS) to Bose-Einstein condensation (BEC) crossover in a two-dimensional Fermi gas with the finite-range interaction by using the Hartree-Fock-Bogoliubov theory.
Expanding the scattering phase shift in terms of the scattering length and effective range, we discuss the effect of the finite-range interaction on the pairing and thermodynamic properties.
By solving the gap equation and the number equation self-consistently, we numerically calculate the effective-range dependence of the pairing gap, chemical potential, and pair size throughout the BCS-BEC crossover.
%We show that we can extract the parameter of the interaction from the peak structure of the pairing gap.
Our results would be useful for further understanding of low-dimensional many-body problems. 
\end{abstract}

\subjectindex{xxxx, xxx}

\maketitle

\section{Introduction}\label{sec:1}
Strongly correlated quantum systems are essential in various contexts of modern physics.
In a fermion system with a weak attractive interaction, it is known that the Bardeen-Cooper-Schrieffer (BCS) state is realized by the formation of Cooper pairs.
If the strength of the attractive interaction becomes stronger, the BCS state turns into the Bose-Einstein condensate (BEC) of tightly bound molecules without any phase transitions~\cite{chen2005bcs,randeria2014crossover,strinati2018bcs,ohashi2020bcs}.
This crossover phenomenon, nowadays called the BCS-BEC crossover, has been proposed originally in electron-hole systems~\cite{Eagles1969}.

After three decades, the BCS-BEC crossover is realized by cold atomic experiments of $^{40}$K and $^{6}$Li~\cite{Regal2004K,Zwierlein2004Li}.
Such cold atom systems, in which the interaction strength can be arbitrarily tuned near the Feshbach resonance~\cite{RevModPhys.82.1225}, have attracted tremendous attention as ideal simulators for other quantum many-body systems, such as superconductors and nuclear matter~\cite{strinati2018bcs,ohashi2020bcs}.
Since the absolute value of the $s$-wave scattering length $a$ can dramatically be enlarged near the Feshbach resonance,
the interparticle interaction can be regarded as a contact-type (i.e., zero-range interaction) and characterized by one parameter $a$.
%When the interaction strength is changed from weak to strong region, the system is changed continuously from the BCS phase to the BEC phase, in which the Cooper pair constructs the tightly bounded bosonic molecule, without any phase transitions.

Recently, the BCS-BEC crossover has been observed not only in cold atom systems but also in condensed matter systems~\cite{kasahara2014field,Shahar2017SC,nakagawa2021gate,PhysRevX.12.011016,liu2022crossover,kim2022evidence}.
In superconducting systems, by tuning the carrier density instead of the strength of interaction, the density-induced BCS-BEC crossover occurs~\cite{nakagawa2021gate}.
This can be understood as the change of the interaction parameters through the Fermi momentum $k_{\rm F}$~\cite{Tajima2022eff3D} [i.e., the dimensionless coupling measure $(k_{\rm F}a)^{-1}$ in three dimensions].
Such a crossover in condensed matter systems is in contrast with that observed in cold atom systems, where $a$ is tuned instead of $k_{\rm F}$.
We note that the density-induced BCS-BEC crossover has been examined in lattice two-color quantum chromodynamics simulation~\cite{10.1093/ptep/ptac137}.
Its three-body analogue has also been discussed~\cite{PhysRevResearch.4.L012021,tajima2023density}.
%While cold atom is limited in dilute Fermi gas due to technicalities of experiment, the career density of superconducting systems can be tuned arbitrarily.
%hence the same phenomenon of BCS-BEC crossover from a unified perspective in different fields.

However, in general the two-body interaction in condensed matter systems, such as superconductors and semiconductors, inevitably involves a finite effective range $R$ in the $s$-wave channel.
It is necessary to discuss how the finite-range interaction affects physical quantities in contrast to cold atom systems.
It is reported that, in the superconducting BCS-BEC crossover, the pairing gap may show a peak structure in the carrier-density dependence, which is not found in cold atom systems~\cite{shi2022density}.
The role of the finite-range interaction for the superconducting dome, that is, the peak structure of the superconducting critical temperature $T_{\rm c}$ in the carrier-density dependence, has also been pointed out in the context of unconventional superconductors~\cite{Langmann2019Dome}.

Moreover, in addition to the effective-range correction, the BCS-BEC-crossover superconductors are observed in a two-dimensional (2D) material~\cite{nakagawa2021gate}.
In 2D systems, stronger correlations can be found compared to the 3D systems because of the reduction of kinetic degrees of freedom, as unconventional superconductors are more easily found in 2D materials than 3D ones.
Remarkably, a two-body bound state can be formed even for an infinitesimally small attraction in 2D~\cite{PhysRevLett.62.981}.
Such a bound-state formation plays a crucial role in the density-induced BCS-BEC crossover.

%Therefore, understanding the behavior of BCS-BEC crossover in two-dimensional systems can lead to the exploration of high-temperature superconductivity by using the cold atom system as a convenient simulator.

In the previous works, the finite-range effects in 2D systems have not been explored systematically yet. 
While the quantum Monte Carlo simulation has been performed with the finite-range interaction, the finite-range dependence has been examined in only the small-range regimes ($0\le k_{\rm F}R\le 0.11$)~\cite{Zielinski2020}.
The effect of the negative effective range has also been examined theoretically~\cite{PhysRevA.96.023619,PhysRevA.101.043607,PhysRevA.102.013313}.
Furthermore, in Ref.~\cite{shi2022density}, the finite-range effect in the 2D superconductor system is considered by fitting to the experimental data but the Hartree-Fock (HF) self-energy contribution, which can be significant in the case with the finite-range interaction~\cite{PhysRevA.97.013601,tajima2019superfluid,PhysRevA.103.063306,Tajima2022eff3D}, has been neglected.
The present authors also studied the finite-range effect in the 2D BCS-BEC crossover by using the Brueckner $G$-matrix approach~\cite{PhysRevA.107.053313}.
However, the effect of the pairing gap has not been taken into account in Ref.~\cite{PhysRevA.107.053313}.

Systematical studies of finite-range effects will also be accessible in future cold atom experiments.
%In the effective-range expansion of the phase shift, the next contributing parameter after the scattering length $a$ is the effective range $R$.
By incorporating the additional process to excited states in the Feshbach resonance mechanism, the two-field optical method has been proposed to arbitrarily tune not only the scattering length but also the effective range~\cite{Haibin2012}.
Furthermore, a similar experiment for controlling the interaction spatially has been performed based on the above proposal~\cite{Arunkmar2019}.

In this paper, we theoretically investigate the effects of the positive effective range in an attractively interacting 2D Fermi gas system by using the Hartree-Fock-Bogoliubov (HFB) theory~\cite{ring2004nuclear}.
The HFB theory is useful to incorporate the finite-range effect and the presence of pairing gap self-consistently with relatively small numerical costs~\cite{PhysRevA.103.063306,Tajima2022eff3D}.
For the validity of the HFB theory,
at least, the mean-field theory should be justified in the weak-coupling ground state corresponding to the BCS region~\cite{PhysRevLett.62.981}.
Moreover, it is known that the mean-field theory can qualitatively describe the BCS-BEC crossover physics at zero temperature, as the information of the two-body bound state is correctly incorporated in the gap equation~\cite{strinati2018bcs,ohashi2020bcs}.
While the $G$-matrix study for the finite-range correction~\cite{PhysRevA.107.053313} does not involve the pairing gap, both the HF self-energy and the pairing gap can be determined self-consistently in the HFB theory.
%For 3D strongly interacting systems, the $G$-matrix method shows qualitatively good results for BCS-BEC crossover and neutron matter~\cite{strinati2018bcs,ohashi2020bcs}.
To understand the finite-range effect on the density-induced BCS-BEC crossover, we numerically calculate the pairing gap and chemical potential, which are directly affected by the effective range through the HF self-energy. To see the microscopic pairing properties, we also examine the pair-correlation length.

This paper is organized as follows.
In Sec.~\ref{sec:2}, we present the theoretical model for the BCS-BEC crossover with an attractive finite-range interaction in 2D.
In Sec.~\ref{sec:3}, we show the numerical results and discuss how the finite-range correction affects the physical quantities such as pairing gap, chemical potential, and pair-correlation length.
In Sec.~\ref{sec:4}, we summarize this paper.

\section{Model}\label{sec:2}
In this section, we introduce the model for the 2D BCS-BEC crossover with the finite-range attractive interaction.
A two-component 2D Fermi gas with the finite-range interaction is considered, where the Hamiltonian is given by
\begin{align}\label{eq:1} H&=\sum_{\bm{k},\sigma}\xi_{\bm{k}}c_{\bm{k}\sigma}^\dag c_{\bm{k}\sigma}
     +\sum_{\bm{k},\bm{k}',\bm{P}}
    V(\bm{k},\bm{k}')
 c_{\bm{k}+\bm{P}/2,\uparrow}^\dag
    c_{-\bm{k}+\bm{P}/2,\downarrow}^\dag
    c_{-\bm{k}'+\bm{P}/2,\downarrow}
    c_{\bm{k}'+\bm{P}/2,\uparrow}.
\end{align}
In Eq.~\eqref{eq:1}, $\xi_{\bm{k}}=k^2/(2m)-\mu$ is the kinetic energy of a fermion with mass $m$ measured from the chemical potential $\mu$, and
$c_{\bm{k}\sigma}^{(\dag)}$ is the annihilation (creation) operator of a fermion with the momentum $\bm{k}$ and the spin $\sigma=\uparrow,\downarrow$. 
We consider the finite-range separable $s$-wave interaction given by
\begin{align}
    V(\bm{k},\bm{k}')=G\Gamma_{k}\Gamma_{k'},
\end{align}
where $G$ is the coupling constant and
\begin{align}
    \Gamma_k=\frac{1}{\sqrt{1+(k/\Lambda)^2}}
\end{align}
is the form factor, which reproduces the relative momentum dependence of the scattering phase shift $\delta_k$ up to $O(k^2)$~\cite{Tajima2019lowEne,Tajima2022eff3D}.
Since we are interested in the attractive interaction, the negative coupling constant $G<0$ is considered.
The momentum scale $\Lambda$ plays a role of the momentum cutoff.
In the following, we show how to relate the model parameters (i.e., $G$ and $\Lambda$) to the 2D scattering length and effective range via the analysis of the two-body $T$-matrix.
In 2D systems, it is known that the two-body bound state always exists with arbitrary small attractive zero-range interaction.
In the case of finite-range interaction,  the $T$-matrix is written by
\begin{align}
    T\left({\bm k},{\bm k'};\omega\right)
    &=G\Gamma_{k}\Gamma_{k'}
    \left[1-G\sum_{\bm{p}}
    \frac{\Gamma_p^2}{\omega_{+}-p^2/m}\right]^{-1},
\end{align}
where $\omega_{+} = \omega + i\delta$ is the two-body energy with an infinitesimally small number $\delta$.
The two-body binding energy $E_{\rm b}$ is obtained from a pole of $T$-matrix as
\begin{align}
    0 &= \frac{4\pi}{mG}+\frac{\log\left(\frac{mE_{\rm b}}{\Lambda^2}\right)}
    {1-\frac{mE_{\rm b}}{\Lambda^2}}.
\end{align}
Also, the scattering length $a$ is given by
\begin{align}
a=\frac{1}{\Lambda}\exp\left(-\frac{2\pi}{mG}\right).
\end{align}
The ratio between the effective range $R$ and $a$ is given by
\begin{align}
\frac{R}{a} = \sqrt{-\frac{4\pi}{mG}\exp\left(\frac{4\pi}{mG}\right)}.
\end{align}
For convenience, we measure the interaction strength and the effective range by using the dimensionless parameters $\log(k_{\rm F}a)$ and $R/a$, where $k_{\rm F}=\sqrt{2\pi \rho}$ is the Fermi momentum.
In the sense of cold atomic physics, $\log(k_{\rm F}a)$ can be tuned by changing $a$ near the Feshbach resonance.
In the density-induced BCS-BEC crossover,
$k_{\rm F}$ and $\log(k_{\rm F}a)$ are changed with the number density $\rho$.
Qualitatively, the dilute BEC (strong-coupling) and dense BCS (weak-coupling) regimes are characterized as $\log(k_{\rm F}a)\lesssim 1$ and $\log(k_{\rm F}a)\gesim 1$, respectively.

Next, the HFB theory is introduced to consider the many-body ground state in the presence
of the nonzero effective range.
To this end, two kinds of the mean-field expectation values are introduced: the pairing gap
\begin{align}
    \Delta(\bm{k})=-\sum_{\bm{k}'}V(\bm{k},\bm{k}')\langle c_{-\bm{k}',\downarrow} c_{\bm{k}',\uparrow} \rangle
\end{align}
and the HF self-energy
\begin{align}
    \Sigma_{\sigma}(\bm{k})=\sum_{\bm{k}'}V\left(\frac{\bm{k}-\bm{k}'}{2},\frac{\bm{k}-\bm{k}'}{2}\right)\langle c_{\bm{k}',\bar{\sigma}}^\dag c_{\bm{k}',\bar{\sigma}}\rangle,
\end{align}
where $\bar{\sigma}$ denotes the opposite spin with respect to $\sigma$.
Since we are interested in the spin-balanced case,
we suppress the spin index as $\Sigma_{\uparrow}(\bm{k})=\Sigma_{\downarrow}(\bm{k})\equiv \Sigma(\bm{k})$.
The resulting mean-field Hamiltonian reads
\begin{align}
\label{eq:H_HFB}
    H_{\rm HFB}
    =&\,\sum_{\bm{k}}\Psi_{\bm{k}}^\dag
    \left[\xi_{\bm{k}}\tau_3+\Sigma(\bm{k})\tau_3-\Delta(\bm{k})\tau_1\right]
    \Psi_{\bm{k}}\cr
    &
    +\sum_{\bm{k}}\Delta(\bm{k})
    \langle c_{\bm{k},\uparrow}^\dag c_{-\bm{k},\downarrow}^\dag \rangle
    -\sum_{\bm{p}}
    \Sigma(\bm{p})
    \langle c_{\bm{p},\uparrow}^\dag c_{\bm{p},\uparrow}\rangle\cr
    &+\sum_{\bm{k}}[\xi_{\bm{k}}+\Sigma(\bm{k})],
\end{align}
where $\tau_{i}$ is the Pauli matrix acting on the Nambu spinor $\Psi_{\bm{k}}=(c_{\bm{k},\uparrow} \ c_{-\bm{k},\downarrow}^\dagger)^{\rm T}$.
After the Bogoiubov transformation, the ground-state energy is written by
\begin{align}\label{ground_state_energy}
    E_{\rm GS}&=
    \sum_{\bm{k}}\Delta(\bm{k})
    \langle c_{\bm{k},\uparrow}^\dag c_{-\bm{k},\downarrow}^\dag \rangle
    -\sum_{\bm{p}}
    \Sigma(\bm{p})    n_{\bm{p}}+\sum_{\bm{k}}\left[\xi_{\bm{k}}+\Sigma(\bm{k})-E_{\bm{k}}\right],
\end{align}
where %$n_{\bm{p}}\equiv\langle c_{\bm{p},\uparrow}^\dag c_{\bm{p},\uparrow}\rangle=\langle c_{\bm{p},\downarrow}^\dag c_{\bm{p},\downarrow}\rangle $
$n_{\bm{p}}=\frac{1}{2}\left(1-\frac{\xi_{\bm{p}}}{E_{\bm{p}}}\right)$
is the momentum distribution and $E_{\bm{k}}=\sqrt{\{\xi_{\bm{k}}+\Sigma(\bm{k})\}^2+|\Delta(\bm{k})|^2}$ is the quasiparticle dispersion.

Moreover, the separable interaction leads to the convinient form of pairing gap as
\begin{align}
    \Delta(\bm{k})=-\Gamma_k G\sum_{\bm{k}'}\Gamma_{k'}
    \langle c_{-\bm{k}',\downarrow} c_{\bm{k}',\uparrow} \rangle
    \equiv \Delta\Gamma_{k},
\end{align}
where the superfluid order parameter
\begin{align}
    \Delta \equiv - G\sum_{\bm{k}'}\Gamma_{k'}
    \langle c_{-\bm{k}',\downarrow} c_{\bm{k}',\uparrow} \rangle
\end{align}
characterizes the magnitude of the pairing gap.
Also, the HF self-energy reads
\begin{align}\label{eq:HFself_energy}
    \Sigma(\bm{k})
    =G\sum_{\bm{k}'}\Gamma_{\frac{|\bm{k}-\bm{k}'|}{2}}^2
    n_{\bm{k}'}.
\end{align}
We note that the momentum dependence of $\Sigma(\bm{k})$ is different from Ref.~\cite{PhysRevA.107.053313}.
While the interaction Hamiltonian is directly replaced by the effective interaction in Ref~\cite{PhysRevA.107.053313},
here Eq.~\eqref{eq:HFself_energy} is derived microscopically under the mean-field approximation.
However, this difference would not change the results qualitatively.

In this way, $E_{\rm GS}$ is further rewritten as
\begin{align}\label{eq:energy_k}
    E_{\rm GS}&=\sum_{\bm{k}}\left[\xi_{\bm{k}}+\Sigma(\bm{k})-E_{\bm{k}}\right]-\frac{|\Delta|^2}{G}
    -\sum_{\bm{p}}
    \Sigma(\bm{p})
    n_{\bm{p}}.
\end{align}
Minimizing $E_{\rm GS}$ with respect to $\Delta$, we obtain the gap equation
\begin{align}\label{eq:gap_equation_HFB}
    1=-G\sum_{\bm{k}}\frac{\Gamma_k^2}{2E_{\bm{k}}}.
\end{align}
To determine $\Delta$ and $\mu$ for a given number density $\rho$ self-consistently,
the gap equation~\eqref{eq:gap_equation_HFB} should be solved with
the number-density equation
\begin{align}\label{eq:number_equation_HFB}
    \rho
    &=\sum_{\bm{k}}
    \left[1-\frac{\xi_{\bm{k}}+\Sigma(\bm{k})}{E_{\bm{k}}}\right].
\end{align}

\begin{figure}[t]
    \centering
    \includegraphics[width=8cm]{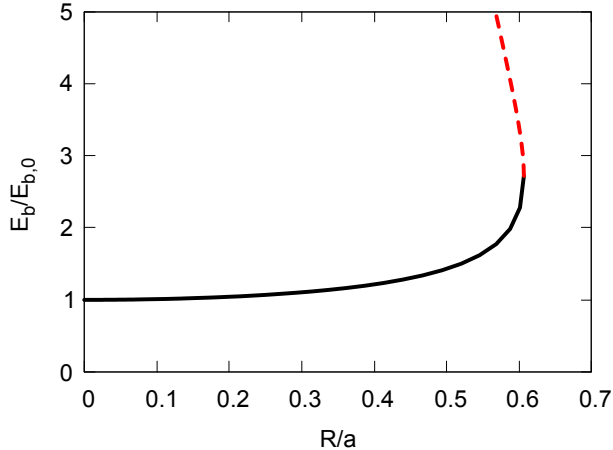}
    \caption{Two-body binding energy $E_{\rm b}$ as a function of $R/a$.
    The energy is normalized by $E_{\rm b,0}=1/ma^2$ at $R/a=0$. 
    The black solid line represents the region directly extended from the result of the contact-type interaction and the red dotted line is the region where a deep bound state is found.
    In this study, the region drawn by the black line is mainly considered. The cutoff $\Lambda$ is chosen as $k_{\rm F}$.
    }
    \label{fig:1}
\end{figure}

In the end of this section,  to be self-contained we review the finite-range effect on a two-body problem in the present model.
Note that the behavior of $E_{\rm b}$ in the present model has already been reported in Ref.~\cite{PhysRevA.107.053313}.
Figure~\ref{fig:1} shows the solution of the two-body binding energy $E_{\rm b}$ as a function of the ratio between the effective range $R$ and the scattering length $a$.
One can check that the zero-range result $E_{\rm b,0}=1/ma^2$ can be obtained in the limit of $R/a\rightarrow 0$.
In this model, $E_{\rm b}$ has two solutions for each $R/a$.
We focus on the solution of smaller $E_{\rm b}$ (solid line in Fig~\ref{fig:1}) because the other solution is unphysically large to discuss the low-energy properties of the present system.
If one increases the effective range $R$, $E_{\rm b}$ is enlarged up to $R/a=e^{-1/2}\simeq0.607$.

\section{Results and discussion}\label{sec:3}

In this section, we present the numerical results of the HFB theory for the 2D BCS-BEC crossover with the finite-range interaction.
The results are obtained by solving Eqs.~\eqref{eq:HFself_energy},  \eqref{eq:gap_equation_HFB}, and \eqref{eq:number_equation_HFB} self-consistently.

\begin{figure*}[t]
\centering
    \includegraphics[width=5.4cm]{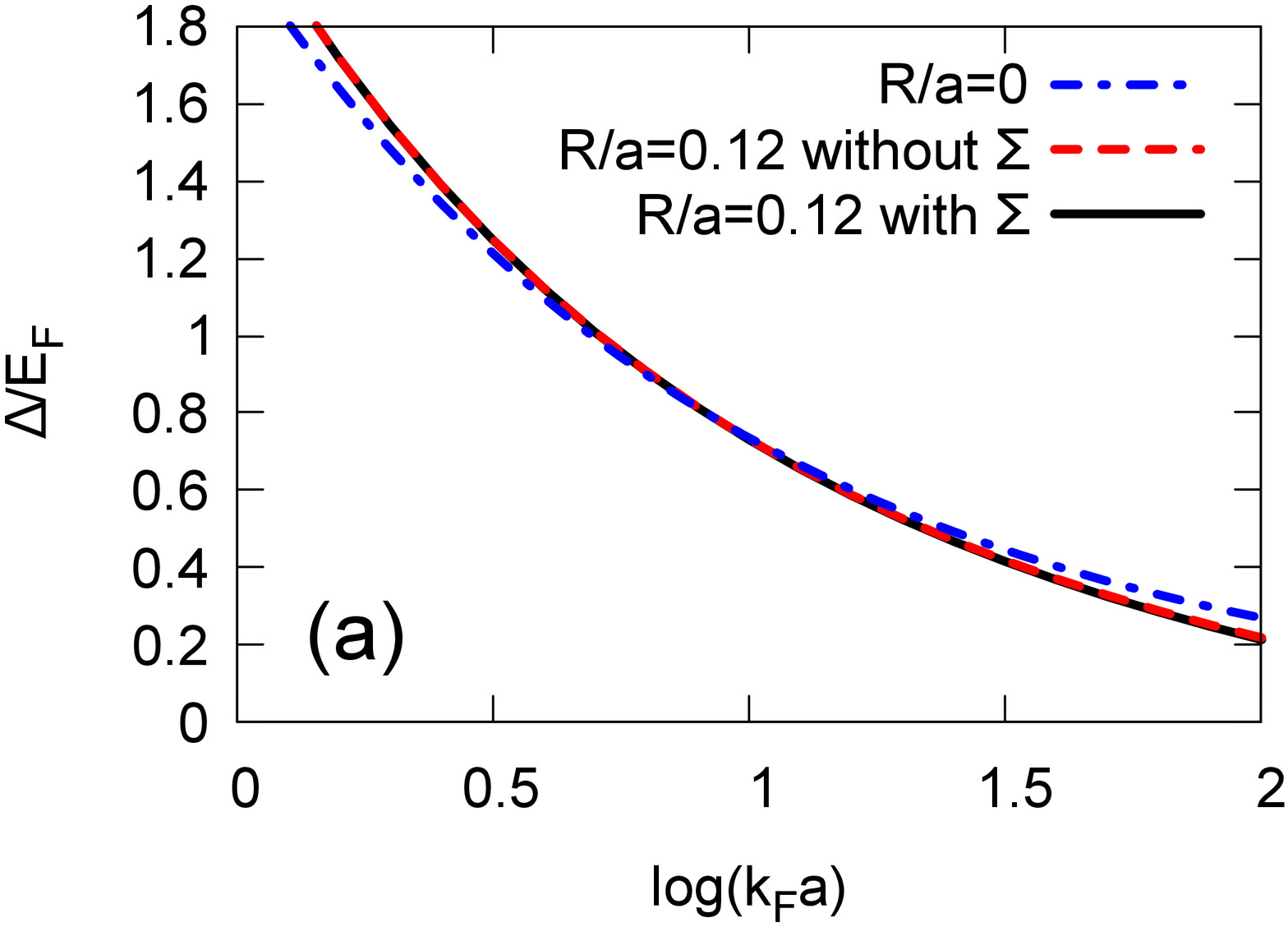}
    \includegraphics[width=5.4cm]{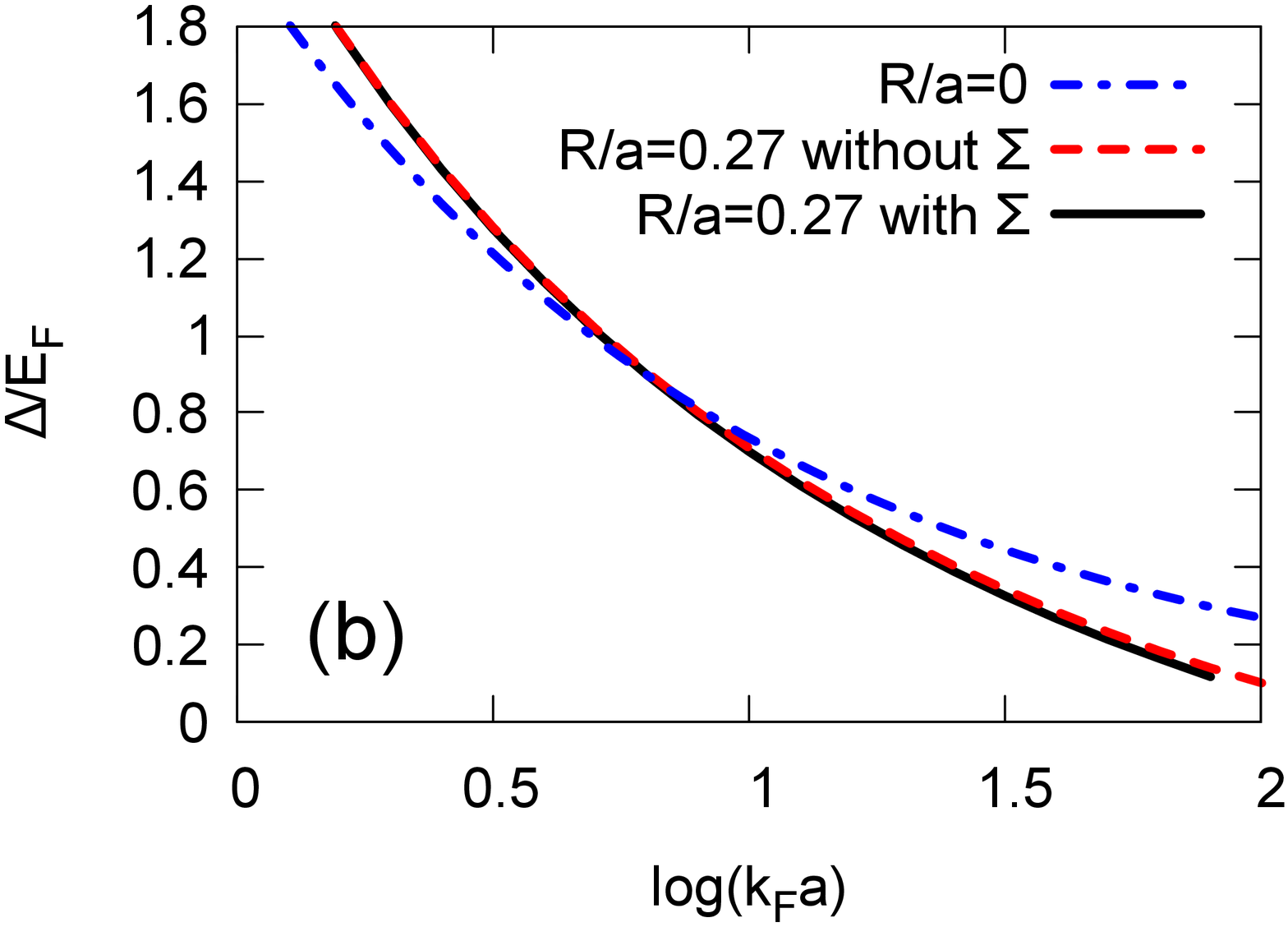}
    \includegraphics[width=5.4cm]{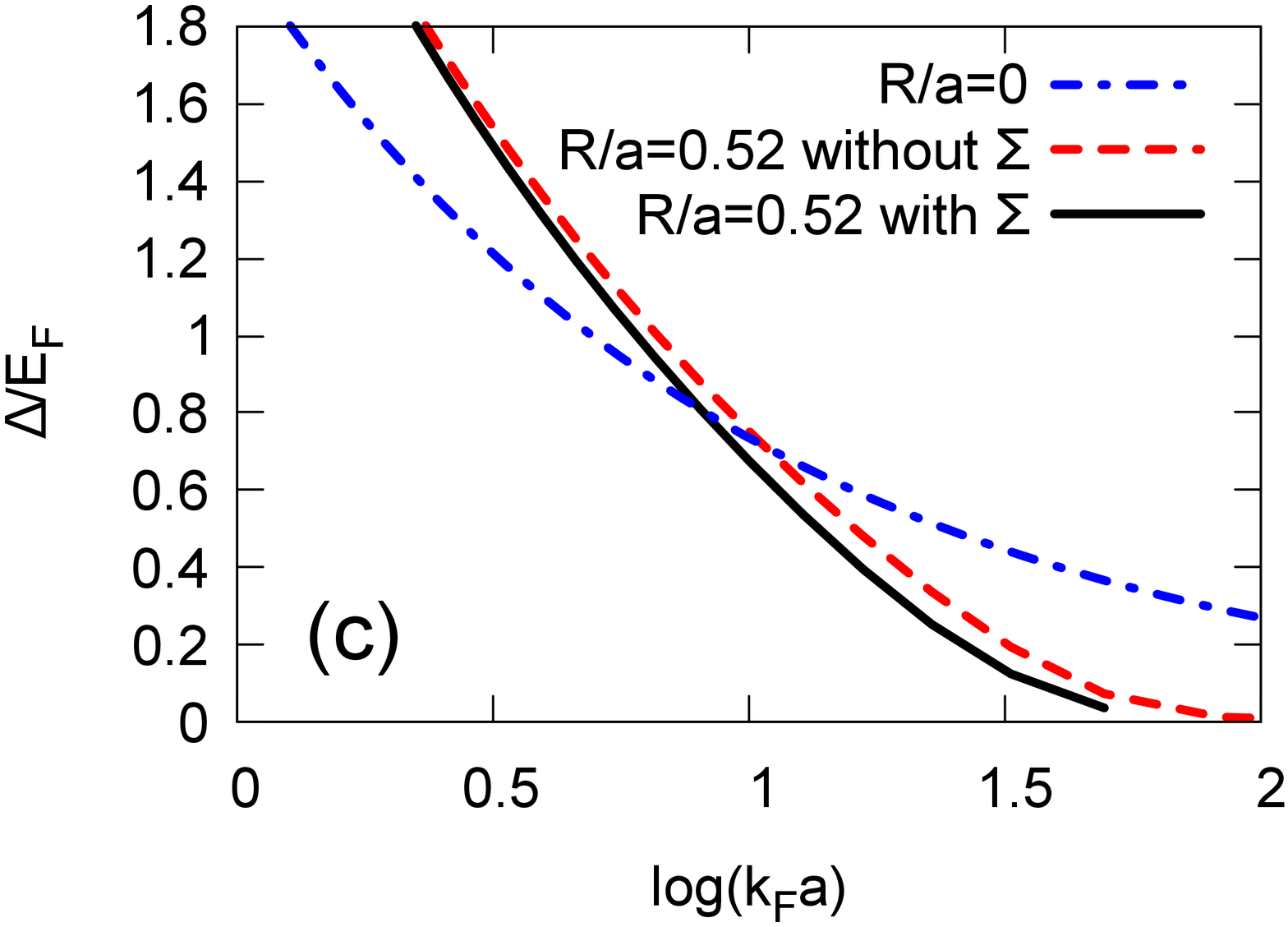}
    \\
    \includegraphics[width=5.4cm]{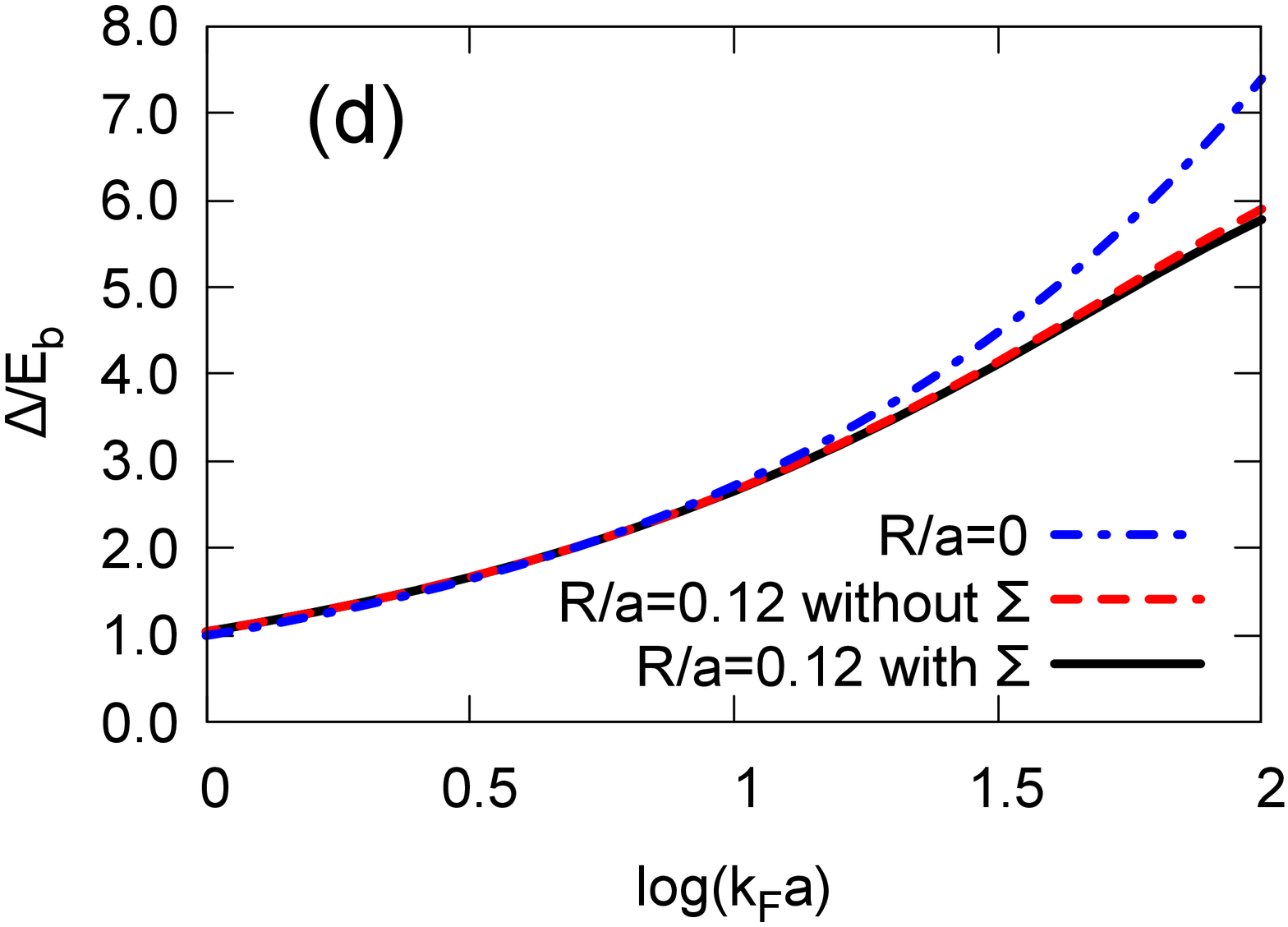}
    \includegraphics[width=5.4cm]{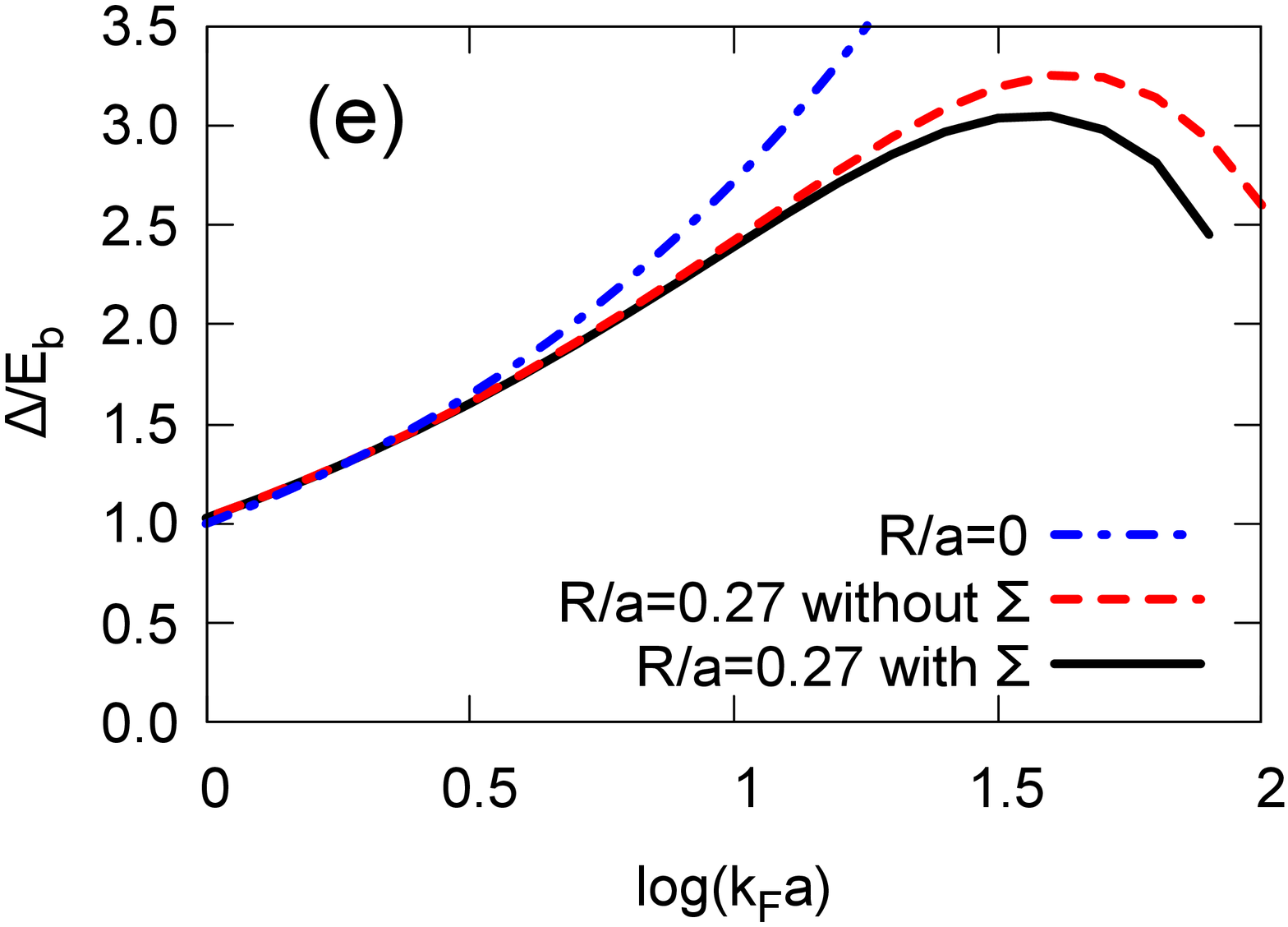}
    \includegraphics[width=5.4cm]{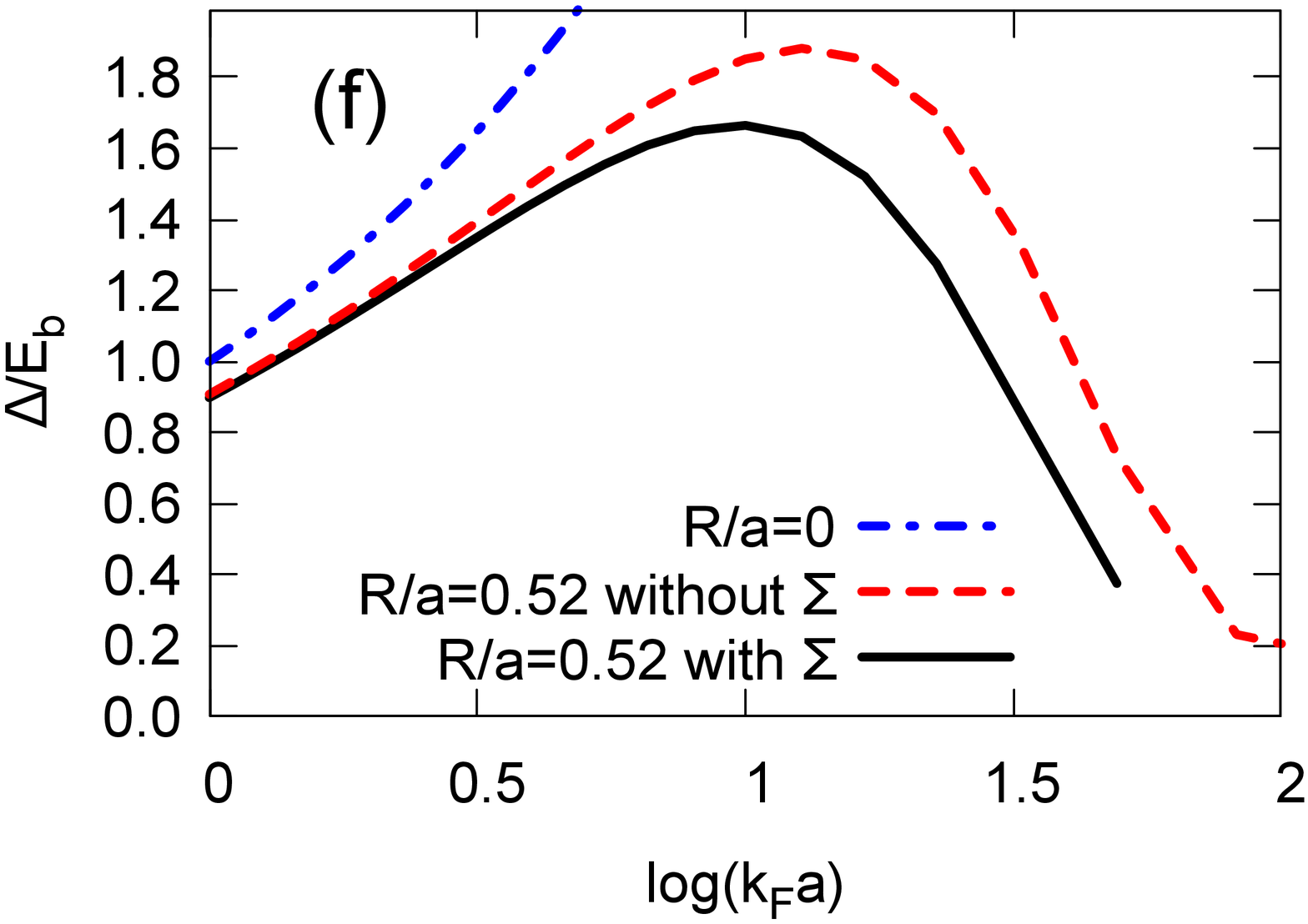}
    \caption{Pairing gap $\Delta$ in the BCS-BEC crossover. In the upper and lower panels, $\Delta$ is normalized by the Fermi energy $E_{\rm F}$ and the binding energy $E_{\rm b}$, respectively. 
    The blue dashed lines are the results calculated with the contact-type interaction. 
    The red dotted lines are the results with the finite-range interaction neglecting the Hartree-Fock self-energy $\Sigma$. 
    The black solid lines are the results with the finite-range interaction including $\Sigma$. 
    The ratios of the effective range $R$ to the scattering length $a$ are chosen as $0.12$, $0.27$ and $0.52$, respectively.}
    \label{fig:2}
\end{figure*}

Figure~\ref{fig:2} shows the pairing order parameter $\Delta$, which characterizes the superfluid order in this system, as a function of the dimensionless coupling parameter $\log(k_{\rm F} a)$ in the 2D BCS-BEC crossover.
The blue dashed lines in Fig.~\ref{fig:2} are the results of contact-type interaction given by~\cite{Randeria1990}
\begin{align}
\Delta(R/a\rightarrow 0) = \sqrt{2E_{\rm F}E_{\rm b,0}},
\end{align}
where $E_{\rm F}=k_{\rm F}^2/2m$ is the Fermi energy. % and $E_{\rm b,0}\equiv E_{\rm b}(R=0)=1/\left(ma^2\right)$ is the two-body binding energy with the zero-range interaction.
$\Delta/E_{\rm F}$ is plotted in panels (a), (b), and (c) with different $R/a$, while $\Delta/E_{\rm b}$ is plotted in panels (d), (e), and (f).
The black solid lines represent the results of the HFB theory with the finite-range interaction.
For comparison, we also show the results with the finite-range interaction without the HF self-energy $\Sigma(\bm{k})$ (the red dashed lines).
This calculation is similar to that used in the previous work for a layered superconductor in the BCS-BEC crossover regime at $T=0$~\cite{shi2022density}.
%Note that some of the black lines stop halfway in the BCS region because of numerical difficulty.

%In the upper panels, the pairing gap is normalized by the Fermi energy $E_{\rm F}$.
In the dilute BEC region $\left(\log\left(k_{\rm F}a\right)\lesssim 1\right)$, the pairing gap is enlarged by the finite-range effect.
This is because the binding energy is enhanced by this effect~\cite{PhysRevA.107.053313} as shown in Fig.~\ref{fig:1}. 
In this regard, fermions can form Cooper pairs more easily than in the case with the contact-type interaction.
On the other hand, in the dense BCS region $\left(\log\left(k_{\rm F}a\right)\gesim 1\right)$,  the formation of Cooper pair is suppressed by the finite-range effect.
Since introducing the finite-range effect is equivalent to introducing the high-momentum cutoff (i.e., $\Lambda$), the pairing order originating from the pairing near the Fermi surface is suppressed when $k_{\rm F}$ is comparable with $\Lambda$. Indeed, the ratio between $k_{\rm F}$ and $\Lambda$ is given by  
\begin{align}
    \frac{k_{\rm F}}{\Lambda}=k_{\rm F}R\sqrt{\frac{m|G|}{4\pi}}.
\end{align}
In this regard, when $R/a$ becomes larger, the suppression of BCS pairing becomes more remarkable.

In the lower panels of Fig.~\ref{fig:2}, the plotted $\Delta$ is normalized by $E_{\rm b}$, which is independent of $\rho$, to clarify the density dependence of $\Delta$.
In Fig.~\ref{fig:2}(d), the finite-range results are similar to the contact-type result, since the parameter $R/a$ is sufficiently small.
$\Delta/E_{\rm b,0}$ with the contact-type coupling increases monotonically as
\begin{align}
\frac{\Delta\left(R/a\rightarrow 0\right)}
{E_{\rm b,0}} 
= k_{\rm F}a.
\end{align}
However, at larger $R$ shown in Figs.~\ref{fig:2}(e) and (f), the peak structure of $\Delta/E_{\rm b}$ can be found with the density dependence [namely, the $\log(k_{\rm F}a)$ dependence] in the finite-range calculations.
This difference clearly manifests the suppression of the BCS-type pairing due to the finite-range correction.
For more details, in Fig.~\ref{fig:2}(e), when $R/a$  increases, the peak structure of the finite-range results can be found at $\log\left(k_{\rm F}a\right) \sim 1.5$.
In Fig.~\ref{fig:2}(f), for larger $R/a$ such a peak structure is more pronounced and shifted toward $\log\left(k_{\rm F}a\right) \sim 1.0$.
In the experiment of the layered superconductor system~\cite{doi:10.1126/science.abb9860}, a similar peak structure has been found as the comparison with the theoretical results has been reported~\cite{shi2022density}.
Such a peak structure is unique to the finite-range interaction and not found in systems with the contact-type interaction.
Therefore, in order to simulate these superconductor systems by using ultracold atom systems, it is necessary to tune the effective range in addition to the scattering length by using e.g., the optical field method~\cite{Haibin2012}.

\begin{figure}[t]
    \centering
    \includegraphics[width=8cm]{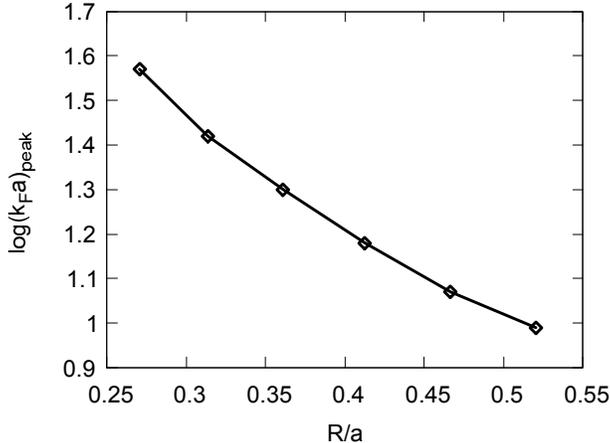}
    \caption{Peak density $\log(k_{\rm F}a)_{\rm peak}$ for the pairing gap $\Delta$ normalized by the binding energy $E_{\rm b}$ (i.e., lower panels of Fig.~\ref{fig:2}.) as a function of $R/a$.
    }
    \label{fig:3}
\end{figure}

Figure~\ref{fig:3} shows the position of the peaks with respect to the ratio $R/a$.
These were obtained by applying the Lagrange interpolation to the numerical data to pick up the maximum value of $\Delta/E_{\rm b}$.
In the limit of $R/a\to0$, the peak position can be at an infinitely large $\log(k_{\rm F}a)$ as the peak structure does not exist in the system with the contact-type interaction.
The peak position $\log(k_{\rm F}a)_{\rm peak}$ tends to decrease monotonically when $R/a$ increases.
This indicates that for larger $R/a$ the peak can be found at lower densities. 
This result would be useful to determine the parameters of interaction from the experiment results with the finite-range interaction, when one tries to qualitatively examine the finite-range properties in condensed matter systems as well as in future cold atom experiments. 

\begin{figure*}[t]
\centering
    \includegraphics[width=5.4cm]{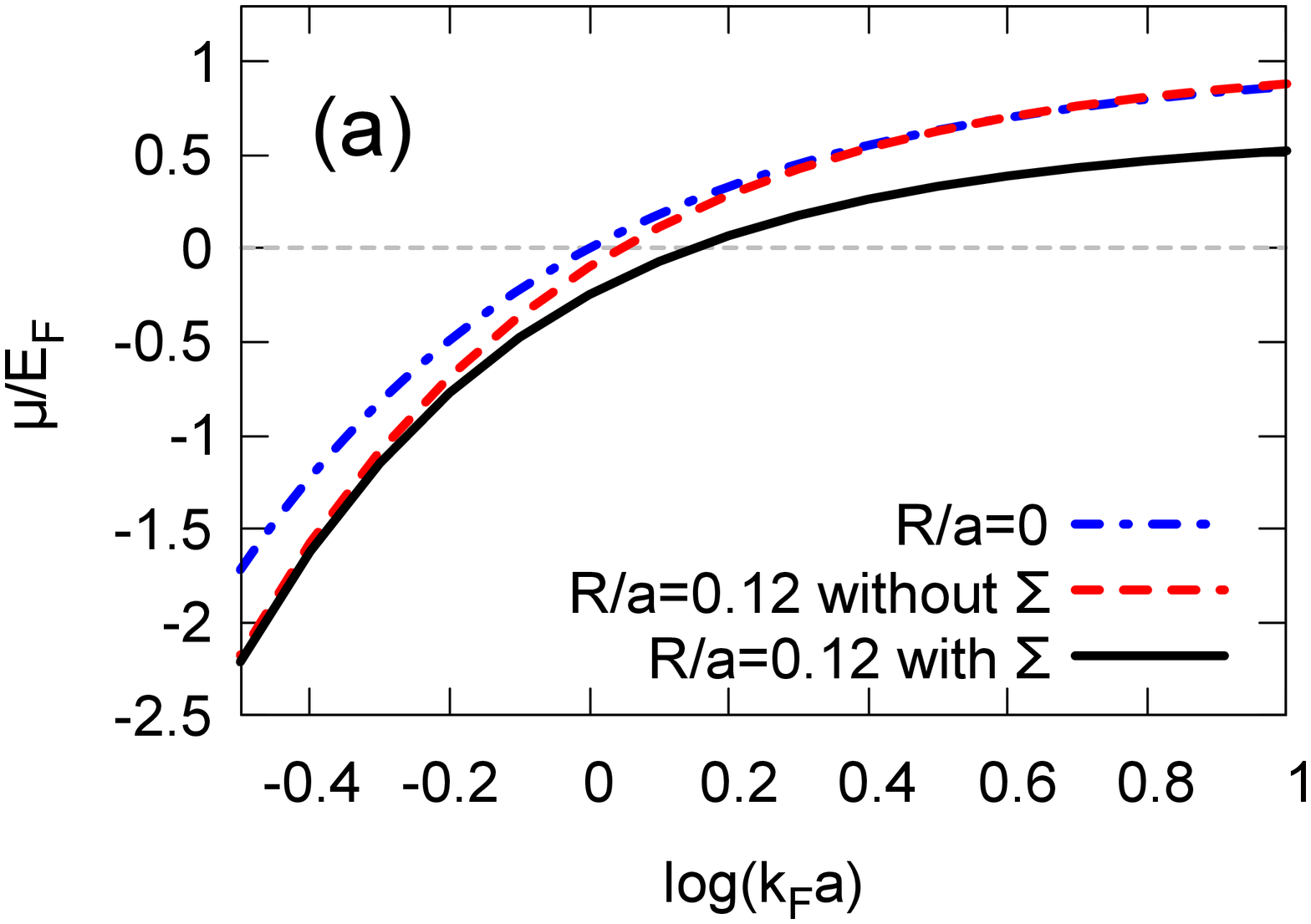}
    \includegraphics[width=5.4cm]{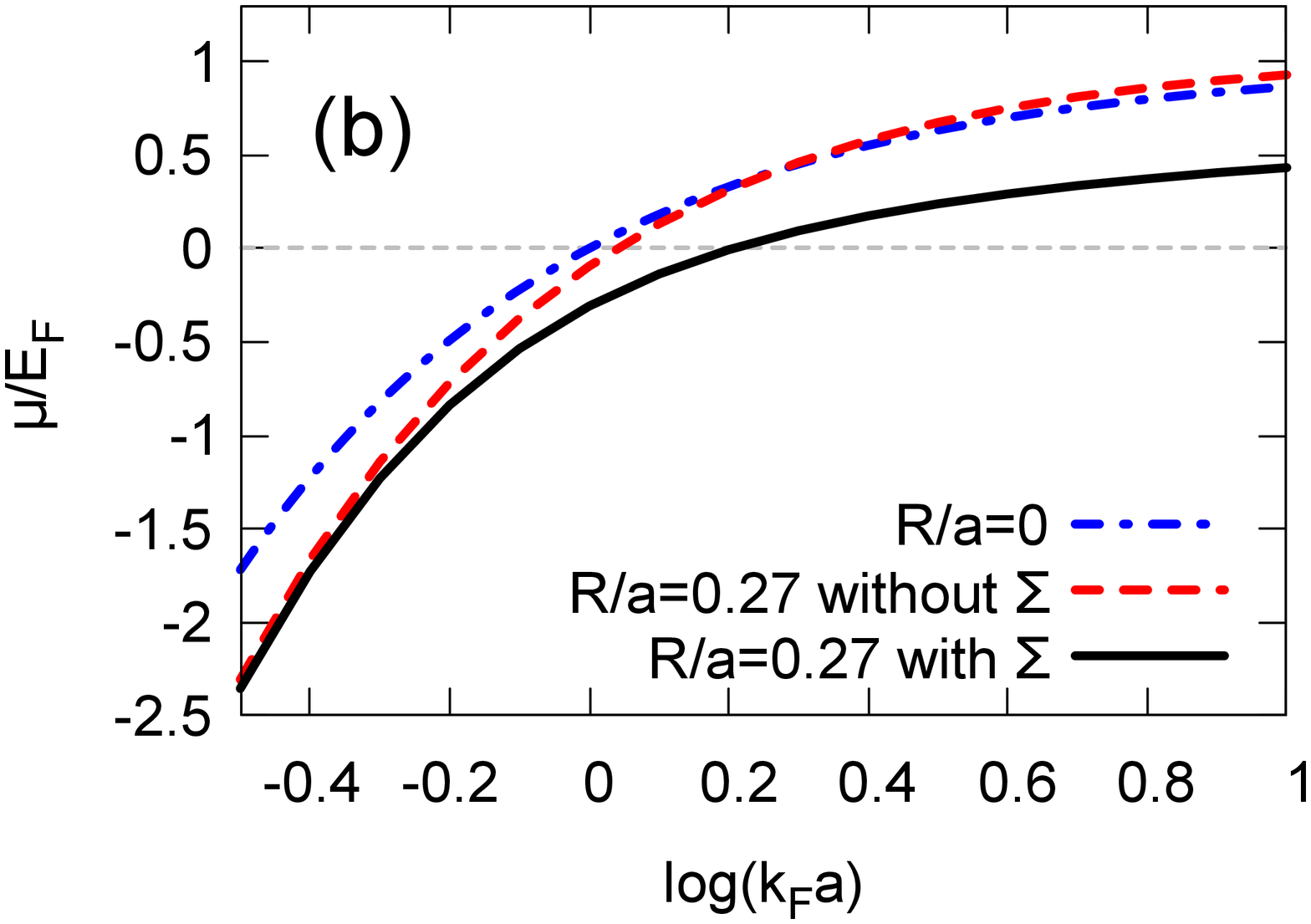}
    \includegraphics[width=5.4cm]{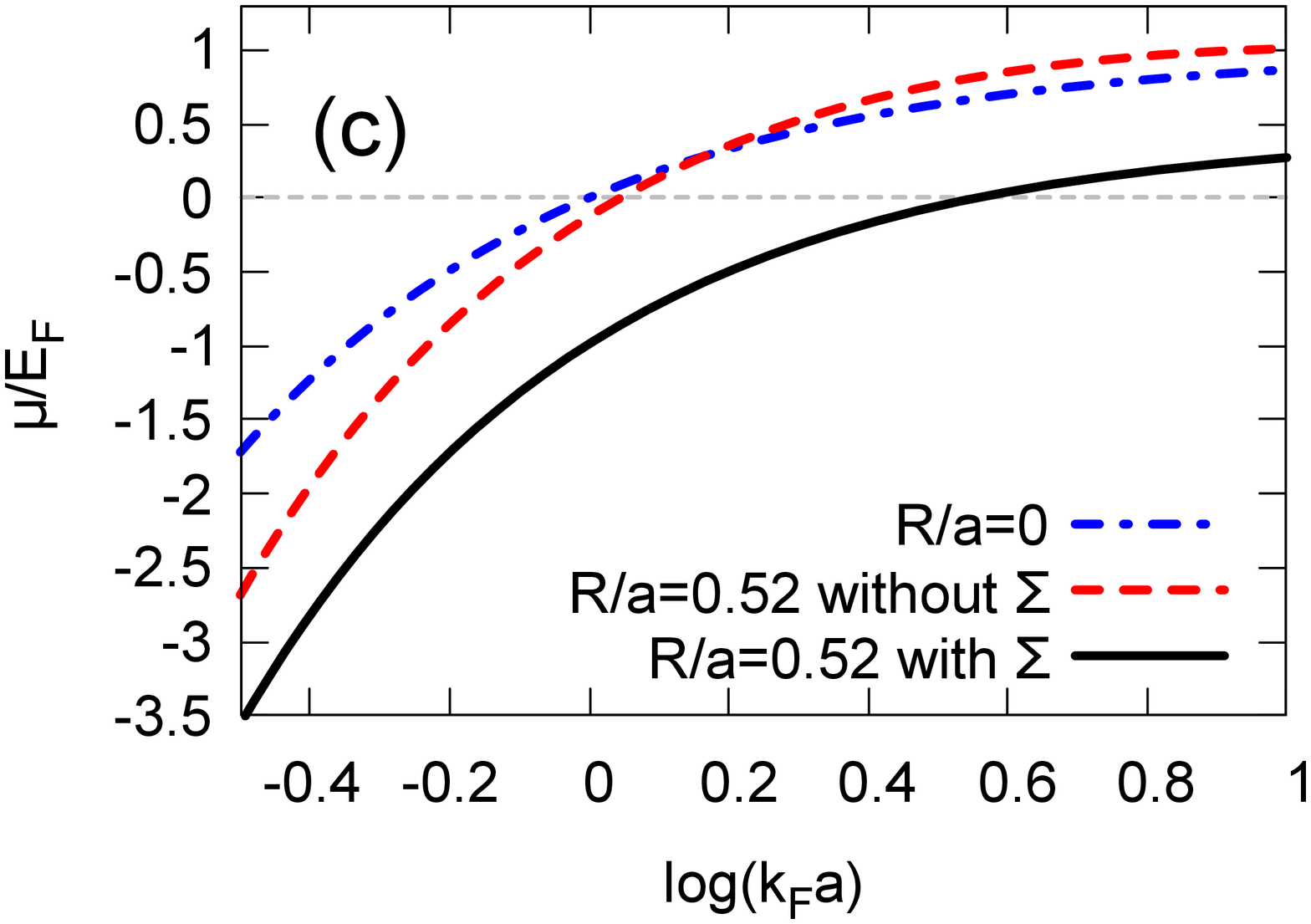}
    \caption{Chemical potential $\mu$ in the unit of Fermi energy $E_{\rm F}$ as a function of $\log\left(k_{\rm F}a\right)$.
    The blue dashed lines are the results with the contact-type interaction.
    The red dotted and black solid lines are the results without and with the HF self-energy, respectively.
    The horizontal dotted lines represent $\mu=0$.
    Here, $R/a$ are chosen as (a) $0.12$, (b) $0.27$, and (c) $0.52$, respectively.}
    \label{fig:4}
\end{figure*}

To see more detailed properties of the density-induced BCS-BEC crossover with the finite range interaction, in Fig.~\ref{fig:4} we show the results of the chemical potential $\mu$ at different effective ranges: (a) $R/a=0.12$, (b) $R/a=0.27$, and (c) $R/a=0.52$.
The blue dashed lines are the results with the contact-type interaction given by
\begin{align}
    \mu = E_{\rm F}-\frac{E_{\rm b,0}}{2}.
\end{align}
While $\mu\simeq E_{\rm F}$ is found in the BCS side,
$\mu\simeq -E_{\rm b,0}/2$ in the BEC side as $\mu$ represents the change of the energy when a single particle is added to the system.
In this regard, $\mu$ is regarded as a thermodynamic quantity well characterizing the BCS-BEC crossover~\cite{ohashi2020bcs}.
One can see that the finite-range effect suppresses $\mu$ in the whole crossover regime.
To understand this suppression of $\mu$ in more detail, we also show the results with the finite-range interaction but without $\Sigma(\bm{k})$ (the red dashed line) for comparison.
It is found that $\mu$ is lowered when $E_{\rm b}$ is enlarged by the finite-range effect.
The reduction of $E_{\rm b}$ is remarkably important in the BEC side as we discussed for the behavior of $\Delta$.
It is also found that $\Sigma(\bm{k})$ (the black solid line) further suppresses $\mu$ compared to the case without $\Sigma(\bm{k})$ in the entire crossover region. 
The effect of $\Sigma(\bm{k})$ is found to be large with increasing $R/a$ but sufficiently small in the dilute BEC side. 
In the dense BCS side, generally $\Sigma(\bm{k})$ gives a significant shift of $\mu$.
This shift of $\mu$ is directly related to $\Sigma(\bm{k}\simeq\bm{0})$ in the quasiparticle dispersion $E_{\bm{k}}=\sqrt{\{\bm{k}^2/2m -\mu+\Sigma(\bm{k})\}^2+|\Delta(\bm{k})|^2}$.
\begin{figure}[t]
    \centering
    \includegraphics[width=8cm]{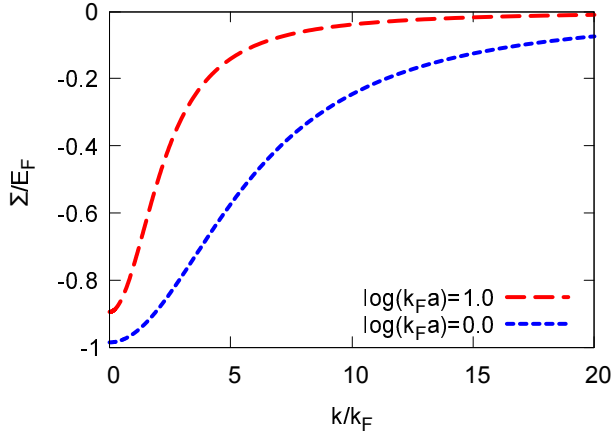}
    \caption{HF self-energy $\Sigma(\bm{k})$ in the unit of Fermi energy $E_{\rm F}$ as a function of the normalized momentum $k/k_{\rm F}$ at $R/a=0.52$.
    The red dashed and blue dotted lines correspond to $\log\left(k_{\rm F}a\right)=1.0$ and $0.0$, respectively. }
    \label{fig:5}
\end{figure}
Figure~\ref{fig:5} shows $\Sigma(\bm{k})$ at $R/a=0.52$,
where $\log(k_{\rm F}a)=1.0$ and $0.0$ are considered.
Indeed, the shift of $\mu$ induced by $\Sigma(\bm{k})$, given by $\Sigma(\bm{k}\simeq\bm{0})\simeq -E_{\rm F}$ at $\log(k_{\rm F}a)=0.0$ and $\Sigma(\bm{k}\simeq\bm{0})\simeq -0.9E_{\rm F}$ at $\log(k_{\rm F}a)=1.0$, are close to the differences between the results with and without  $\Sigma(\bm{k})$ in Fig.~\ref{fig:4}(c).
We note that $\Sigma(|\bm{k}|\simeq k_{\rm F})$ is also similar to $\Sigma(\bm{k}\simeq 0)$ in this regime.
This result indicates that the momentum-independent Hartree shift used in Refs.~\cite{PhysRevA.97.013601,tajima2019superfluid,PhysRevC.82.024911} gives a reasonable approximation.
The momentum dependence of $\Sigma(\bm{k})$ in Fig.~\ref{fig:5} is characterized by $\Gamma_{k}$ and hence $\Lambda$.
At low energy, the momentum dependence of $\Sigma(\bm{k})$ may lead to the effective mass correction~\cite{tajima2023polaronic}.

The zero-crossing point of $\mu$, which indicates the interaction strength where the underlying Fermi surface is depleted by the pair formation, has conveniently been regarded as the crossover boundary between the BCS and BEC sides~\cite{Levinsen2014} whereas there are no distinct phase boundaries between them.
While the zero-crossing point of $\mu$ can be found for arbitrary $R/a$ as shown in Fig.~\ref{fig:4},
such a point is quantitatively shifted by the finite-range correction through $\Sigma(\bm{k})$. 
In contrast to the zero-range case, where the HF self-energy is trivially zero and $\mu$ is mainly reduced by the pair formation, we need to consider two possibilities of the reduced $\mu$ in the case with the finite-range interaction, that is, the pair formation and the HF self-energy shift.
In other words, at large $R/a$, $\mu$ can be strongly suppressed by $\Sigma(\bm{k})$ even with the small pairing effect.
In this regard, one needs to carefully examine $\mu$ when trying to use $\mu$ the measure of the BCS-BEC crossover with the finite-range interaction.

\begin{figure}[t]
    \centering
    \includegraphics[width=8cm]{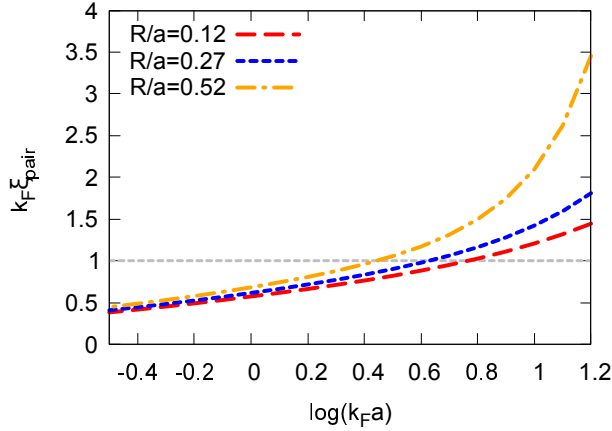}
    \caption{Pair-correlation length $k_{\rm F}\xi_{\rm pair}$ as a function of $\log\left(k_{\rm F}a\right)$, where $R/a=0.12$, $0.27$, and $0.52$ are employed.
    The horizontal dotted line represents $k_{\rm F}\xi_{\rm pair} = 1$, where the pair size is comparable with the mean interparticle distance qualitatively given by $k_{\rm F}^{-1}$.}
    \label{fig:6}
\end{figure}

Finally, to further investigate microscopic properties of Cooper pairs in the present system, 
in Fig.~\ref{fig:6} the result of the pair-correlation length $\xi_{\rm pair}$ is plotted, where
$\xi_{\rm pair}$ is
defined by~\cite{Pistolesi1994}
\begin{align}
    \xi_{\rm pair}^2=\frac{\sum_{\bm{k}}|\nabla_k\phi(\bm{k})|^2}{\sum_{\bm{k}}|\phi(\bm{k})|^2}.
\end{align}
The pair-correlation length is also regarded as a useful quantity to characterize the BCS-BEC crossover~\cite{strinati2018bcs,ohashi2020bcs}. 
In the dilute BEC regime, fermions form tightly bound bosonic molecules and hence $\xi_{\rm pair}$ becomes smaller.
In the dense BCS regime, loosely-bound Cooper pairs are formed and their sizes are typically larger than the mean interparticle distance given by $k_{\rm F}^{-1}$. 
This behavior is not changed significantly by the finite effective range correction.
Entirely, the finite-range effect enlarge $\xi_{\rm pair}$ in the crossover regime.
In particular, in the dense BCS side ($\log(k_{\rm F}a)\gesim 1$), $\xi_{\rm pair}$ is dramatically enlarged by the effective range correction, indicating the suppression of the BCS-type pairing by the finite-range effect. 
While $\mu$ is suppressed by the finite-range effect through two mechanisms, that is, the Cooper pairing and the HF self-energy shift,
$\xi_{\rm pair}$ monotonically increases with $R/a$ and is more directly related to the Cooper pairing effect.
One may expect that the enhanced $\xi_{\rm pair}$ is also related to the tremendous suppression of $\Delta$ and the resulting peak structure in the density dependence of $\Delta/E_{\rm b}$ in the dense BCS regime as shown in Fig.~\ref{fig:2}.
In this regard, the finite-range effect can be found in a different way through the different physical quantities, and $\xi_{\rm pair}$ would be more convenient than $\mu$ to characterize the density-induced BCS-BEC crossover with the finite-range interaction.
In addition, $k_{\rm F}\xi_{\rm pair}=1$ can be used as a crossover boundary between the BCS and BEC regimes.
In such a viewpoint, 
the density with $k_{\rm F}\xi_{\rm pair}=1$ is shifted toward the lower densities when $R/a$ increases.
This result again indicates the suppression of the Cooper pairing near the Fermi surface.
This is in contrast to the shift of zero-crossing point of $\mu$ toward higher densities with increasing $R/a$.
We note that in the HFB framework the effective Fermi surface locates at the shifted chemical potential $\mu^*=\mu-\Sigma(|\bm{k}|\simeq k_{\rm F})$~\cite{PhysRevA.93.013610,PhysRevA.95.043625} and therefore negative $\mu$ does not immediately mean the disappearance of the Fermi surface as we find that the zero-range result and the finite-range result without $\Sigma(\bm{k})$ in Fig.~\ref{fig:4} are close to each other in the dense BCS regime.
In this way, one can understand that $k_{\rm F}\xi_{\rm pair}=1$ is a more appropriate indicator of the BCS-BEC crossover than $\mu=0$.

We also remark that the cluster size (corresponding to $\xi_{\rm pair}$ in this paper) is highly important to understand the microscopic properties of the density-induced hadron-quark crossover~\cite{10.1093/ptep/ptac137,PhysRevD.105.076001}.
Indeed, the overlapped three-body state, which is larger than the interparticle distance, can be anticipated in the dense regime~\cite{PhysRevResearch.4.L012021,tajima2023density}.
Our study on the role of the finite effective range for the pair size would be useful for further extensions to other crossover phenomena.

%This result indicates that finite range effect has a different tendencies in each quantity. The reduction of chemical potential would be two ways, that is, the formation of two-body bound state and HF self-energy as discussed above.

\section{Summary and perspectives}\label{sec:4}

In this paper, we have theoretically investigated the finite-range effect in the 2D Fermi gas system throughout the BCS-BEC crossover by using the Hartree-Fock-Bogoliubov theory. 
Using the finite-range separable interaction, 
we have numerically solved the particle number equation and the gap equation self-consistently.
The momentum-dependent HF self-energy, which were ignored in previous studies, has been considered in the numerical calculation.
The finite-range effects for the pairing gap, the chemical potential, and the pair size are studied systematically.

In particular, the finite-range effect works on the pairing gap in different ways in the BCS and BEC sides, respectively.
While the pairing gap is enhanced by the finite-range effect (through the enhancement of the two-body binding energy) in the BEC side, it is suppressed by the finite-range effect in the BCS side because the effective pairing interaction near the Fermi surface is suppressed by the cutoff associated with the effective range.
Furthermore, the maximal behavior of $\Delta$ normalized by the density-independent scale is identified as density dependent and the peak density is plotted as a function of the effective range.
For the suppression of chemical potential $\mu$ by the finite-range effect, there are two mechanisms, that is, the enhanced pairing correlations and the HF self-energy shift.
In this regard, one needs to carefully examine the effective-range correction to understand the behavior of $\mu$ in the density-induced BCS-BEC crossover.
Finally, we have examined the finite-range effect on the pair correlation length throughout the density-induced BCS-BEC crossover.
The pair size is found to be monotonically enlarged by the finite effective range in the whole crossover region and gives a useful measure for the BCS-BEC crossover from a microscopic viewpoint.

For future perspectives, for further understanding of the connection between clean cold atom systems and other condensed matter systems, it is important to generalize our approach to more realistic interaction model, such as the Rytova-Keldysh potential~\cite{rytova2018screened,keldysh1979coulomb}. 
To obtain more quantitative results, the HF self-energy can be further renormalized by using the Brueckner $G$-matrix~\cite{PhysRevA.107.053313}, where the repeated scattering process is effectively included.
Also, while we have focused on the ground-state properties at zero temperature, it is an interesting future work to examine how the Berezinskii-Kosterlitz-Thouless transition is modified by the finite-range effect and the associated HF self-energy shift~\cite{berezinskii1972destruction,Kosterlitz_1973,Kosterlitz_1974}.

\section*{Acknowledgements}
H.S. was supported by RIKEN Junior Research Associate Program.
H.T. acknowledges the JSPS Grants-in-Aid for Scientific Research under Grant Nos.~18H05406, 22K13981, and 22H01158.
H.L. acknowledges the JSPS Grant-in-Aid for Early-Career Scientists under Grant No.~18K13549, the JSPS Grant-in-Aid for Scientific Research (S) under Grant No.~20H05648, and the RIKEN Pioneering Project: Evolution of Matter in the Universe.

\bibliographystyle{ptephy}
\bibliography{reference.bib}

\end{document}